Effect of high annealing temperature on giant tunnel magnetoresistance ratio of CoFeB/MgO/CoFeB magnetic tunnel junctions

J. Hayakawa[1,2], S. Ikeda[2], Y. M. Lee[2], F. Matsukura[2], and H. Ohno[2]

*1. Hitachi, Ltd., Advanced Research Laboratory, 1-280 Higashi-koigakubo, Kokubunji-shi, Tokyo 185-8601, Japan*

*2. Laboratory for Nanoelectronics and Spintronics, Research Institute of Electrical Communication, Tohoku University, 2-1-1 Katahira, Aoba-ku, Sendai 980-8577, Japan*

We report tunnel magnetoresistance (TMR) ratios as high as 472% at room temperature and 804% at 5 K in pseudo-spin valve (SV) CoFeB/MgO/CoFeB magnetic tunnel junctions (MTJs) annealed at 450$^{\circ}$C, which is approaching the theoretically predicted value. By contrast, the TMR ratios for exchange-biased (EB) SV MTJs with a MnIr antiferromagnetic layer are found to drop when they are annealed at 450$^{\circ}$C. Energy dispersive X-ray analysis shows that annealing at 450$^{\circ}$C induces interdiffusion of Mn and Ru atoms into the MgO barrier and ferromagnetic layers in EB-SV MTJs. Mechanisms behind the different annealing behavior are discussed.

E-mail: j-hayaka@rd.hitachi.co.jp

Soon after tunnel magnetoresistance (TMR) ratios over 1000% were theoretically predicted in the magnetic tunnel junctions (MTJs) with a crystalline (100) Fe/MgO/Fe structure by preferential tunneling of "half-metallic" $\Delta_1$ symmetry states[1,2], TMR ratios ranging from 80 – 355% at room temperature (RT) was experimentally demonstrated for fully (100) oriented epitaxial Fe/MgO/Fe MTJs[3,4] and sputter-deposited highly (100) oriented Fe(CoB)/MgO/Fe(CoB) MTJs[6-8]. The highest TMR ratio so far reported reaches 361 % for a sputter-deposited CoFeB/MgO/CoFeB structure[9] and 410 % for an epitaxial Co/MgO/Co structure.[10] Realizing high TMR ratios by sputtering is of prime importance from a technology point of view, because sputtering is the preferred and established method for industrial applications. Our recent study showed that TMR ratios of 361% at RT and 578% at low temperature can be achieved for standard exchange-biased spin valve (EB-SV) MTJs having sputtered CoFeB/MgO/CoFeB with an underlying synthetic ferrimagnetic (SyF) layer pinned magnetically by an antiferromagnetic (AF) layer beneath. We found that TMR ratios started to decrease when annealing temperatures ($T_a$) exceeded 425°C.[8,9] Several mechanisms were suggested in the past as to the reasons for the decreased TMR ratios at high $T_a$; among these, Mn diffusion from the AF layer into the ferromagnetic layer and towards the barrier insulator is the most suspected in Al-O barrier MTJs.[11,12]

In this letter, we fabricated pseudo SV (P-SV) MTJs that do not have Mn-based AF and SyF pinned layers with (100) oriented CoFeB/MgO/CoFeB by sputtering and annealed them at 250 – 475°C. We then compared the resultant TMR ratios with those of EB-SV MTJs with AF (MnIr) and SyF pinned layers. In the following, we mainly focus on MTJs having $Co_{40}Fe_{40}B_{20}$ electrodes and in the end briefly report the results of MTJs with Fe-rich electrodes.

Figure 1 shows schematic diagrams of the fabricated P-SV (a) and EB-SV MTJs (b). The MTJ films were deposited on $SiO_2$/Si substrates using rf magnetron sputtering with a base pressure of $10^{-9}$ Torr. The order of the film layers starting from the substrate side is as follows: Ta(5) / Ru(20) / Ta(5) / CoFeB(4.3) / MgO (1.5) / CoFeB(4) / Ta(5) / Ru(5) for the P-SV MTJ; and Ta(5) / Ru(50) / Ta(5) / NiFe(5) / MnIr(8)/ CoFe (2.5)/ Ru(0.8)/ CoFeB(3) / MgO (1.5) / CoFeB(3) / Ta(5) / Ru(5) for the EB-SV MTJ. The numbers in parentheses indicate the thickness of the layers in nm. We use CoFeB to represent a $Co_{40}Fe_{40}B_{20}$ or $Co_{20}Fe_{60}B_{20}$ alloy. The MgO layer was formed using an MgO target at a pressure of 10 mTorr in an Ar atmosphere. The MTJs were annealed at $T_a$ from 250 to 475°C for 1 h in a vacuum of $10^{-6}$ Torr under a magnetic field of 4 kOe. All junctions, from 0.8 x 0.8 μm$^2$ to 0.8 x 5.6 μm$^2$, were fabricated using a conventional photolithography process. The electrical properties of the MTJs were measured at RT and at 5 K using a four-probe method with a dc bias and a magnetic field of up to 1 kOe. TMR ratios at 5 K were measured in a field-cooled state (-2 kOe).

The major TMR loops for a typical P-SV MTJ measured with a dc bias voltage of 1mV at RT and 5 K are shown in the top graph of Fig. 2. The obtained TMR ratios of the $Co_{40}Fe_{40}B_{20}$ electrode sample at 450°C (Fig. 2 (a)) were found to be 450% at RT and 747% at 5 K (we discuss about Fig. 2 (b) later in this paper). Using Julliere's formula,[13] these TMR ratios correspond to effective tunneling spin-polarizations of 0.83 at RT and 0.89 at 5 K. This TMR ratio at RT is considerably higher than that of EB-SV MTJs; 361% at optimum $T_a$ of 425°C.[9] The temperature dependence of the TMR ratio and the resistances in parallel (P) and anti-parallel (AP) configurations are plotted in Fig. 2 (c). The resistance in the AP configuration increases as temperature decreases, while that in the P configuration remains almost constant. Hence, the temperature dependence of the TMR ratio is caused by the temperature dependence of the resistance in the AP configuration.

Next, we show the $T_a$ dependence of TMR ratios and compare the results with those of EB-SV MTJs. Figure 3 (a) shows TMR ratios measured at RT as a function of $T_a$ for P-SV MTJs (open circles) and EB-SV MTJs (solid circles). The two curves are virtually the same at $T_a < 400$°C. As $T_a$ exceeds 400°C, the TMR ratio for P-SV MTJs continues to increase up to 450% at $T_a = 450^oC$ and then decrease slightly to 440% at 475$^o$C. By contrast, the TMR ratios for EB-SV MTJs start to decrease at $T_a$=425$^o$C and continue decreasing to values as low as 160% at $T_a$=450$^o$C. Parenthetically, the difference between the results on EB-SV MTJs in ref. 8 and the EB-SV MTJs reported here is due to the thickness of the Ru layer (see ref. 8 for details).

For EB-SV MTJs having an AF layer containing Mn atoms, several mechanisms have been suggested for the drop in TMR ratios at high $T_a$. These include, Mn diffusion from the AF into the ferromagnetic (F) layer and towards the barrier[11,12], small exchange bias field $(H_{ex})$[9], and crystallization to face-centered cubic (fcc) structure/ (110) oriented body centered cubic (bcc) in the ferromagnetic layers.[9,14] We first consider the effect of $H_{ex}$, which is plotted as a function of $T_a$ in Fig. 3 (b). $H_{ex}$ is defined as the external magnetic field at which the TMR ratio becomes half of its maximum value as shown in insets of Fig. 3 (b), where the TMR loops of the EB-SV MTJs with $T_a$=350$^o$C (top) and $T_a$=450$^o$C (bottom) are shown, respectively. Here, a positive $H_{ex}$ is observed when $T_a$=350$^o$C, whereas it became negative when $T_a$ is increased to 450$^o$C. Positive (Negative) $H_{ex}$ shows that the magnetizations between CoFeB and CoFe in the SyF pinned layer are antiferromagnetically (ferromagnetically) coupled. From Fig. 3(b), $H_{ex}$ is found to decrease as $T_a$ increases, and finally changing the sign from positive to negative value. This reversed sign of $H_{ex}$, *i.e.* the change from AF coupling to F coupling in the SyF pinned layer, is most likely due to the reduction of the Ru layer due to diffusion as $T_a$ increases. In our previous work, we found that a weak $H_{ex}$ results in incomplete AP configuration between the two CoFeB layers and in decreased TMR ratios, whereas when $|H_{ex}|$ is large enough to produce a complete AP configuration, high TMR ratio is maintained.[9] The result of $T_a$=450°C in the inset of Fig. 3 (b) shows that a full AP configuration is realized, indicating that the reduction and subsequent sign change of $H_{ex}$ is not responsible for the observed reduction of the TMR ratio.

To investigate the effect of the crystalline structure and diffusion of constituent elements, especially of Mn and Ru, on the TMR ratio, we carried out cross-sectional high resolution transmission electron microscopy (HRTEM) analysis with energy dispersive X-ray (EDX) spectroscopy of the MTJ structures using a 1-nm diameter spot. Figure 4 shows HRTEM images in the region of the CoFeB/MgO/CoFeB interface for a P-SV MTJ annealed at 450°C (a), an EB-SV MTJ annealed at 375°C (b), and an EB-SV annealed at 450°C (c). Three features can be pointed out: (1) For all MTJs, the MgO barriers have a highly (100) oriented NaCl structure and are uniform, and the CoFeB ferromagnetic electrodes, initially amorphous (not shown)[7], have crystallized to a highly oriented (100) bcc structure. (2) In Figs. 4 (a) and (c), MgO barriers are atomically flat and the CoFeB/MgO interfaces are atomically sharp. (3) For EB-SV MTJs (Figs. 4 (b) and (c)), annealing at higher temperatures results in sharper CoFeB/MgO interfaces. Even though the sharpness of the interface improves by high temperature annealing, the TMR ratio decreases at high $T_a$ (Fig. 3(a)) in EB-SV MTJs. This signifies that the crystalline properties are not likely to be the main reason for the drop of TMR ratio at $T_a$ over 425°C.

Figure 5 shows the EDX spectra ranging from Ru to Mn for the six points (a) – (f) indicated in Fig. 4 (b) and (c). A clear Mn peak exists in the SyF layer and even in the MgO barrier already at $T_a$=375°C (Figs. 5 (b) and (c)), indicating that the Mn atoms diffuse towards the MgO barrier. At this $T_a$, the TMR ratio of EB-SV MTJ is virtually the same as that of P-SV MTJs (see Fig. 3). Therefore, the Mn diffusion of this level into the MgO barrier and SyF layer does not affect the TMR ratio, consistent with the results reported recently on amorphous AlO barrier MTJs.[15] By contrast, the peak intensities of Mn atoms after annealing at 450°C are found to be much enhanced; Mn to Mg intensity ratio is 0.03 for $T_a$=375°C and 0.10 for $T_a$=450°C. Furthermore, notable peaks of Ru in the MgO barrier (Fig.5 (e)) is observed, which are not present at 375°C (Fig.5 (b)), indicating that Ru diffuses into the MgO barrier as well as Mn at $T_a$=450°C.

The peaks of Mn and Ru are virtually absent in the top CoFeB layers for both EB-MTJs annealed at 375°C and at 450°C (Fig. 5 (a) and (d)). Although not conclusive, interdiffusion of high-level of Mn and/or Ru atoms may be responsible for the drop in TMR ratios at $T_a$=425°C in EB-SV MTJs.

Finally, we have examined the effect of electrode composition by replacing $Co_{40}Fe_{40}B_{20}$ with Fe-rich $Co_{20}Fe_{60}B_{20}$ in P-SV MTJs. After annealing at 450°C, the MTJs showed TMR ratios as high as 472% at RT and 804% at 5 K (Fig. 2 (b)), approaching the theoretically predicted value of 1000%. This further increase of the TMR ratio is believed to be a compromised result of two competing tendencies: Fe tends to be bcc (favorable for a high TMR ratio) but have reduced spin polarization compared to Co. Co has a tendency to be fcc/hexagonal close packed (hcp) that is not preferable for realizing a high TMR ratio. Work is in progress to clarify the role of electrode composition on the TMR ratio.

In conclusion, we observed TMR ratios as high as 472% at RT and 804% at 5 K in P-SV CoFeB/MgO/CoFeB MTJs annealed at 450°C. By contrast, the TMR ratios for EB-SV MTJs with a MnIr AF layer were found to drop at $T_a$ over 425°C. EDX analysis showed that Mn diffuses into the MgO barrier of EB-SV MTJs annealed at 375°C, while the TMR ratio is virtually the same as that for P-SV MTJ. Increasing annealing temperature further to 450°C induced diffusion of Ru and Mn into the MgO barrier, which may be the reason for the drop in TMR ratios in EB-SV MTJs.

This work was supported by the IT-program of the Research Revolution 2002 (RR2002): "Development of Universal Low-power Spin Memory", Ministry of Education, Culture, Sports, Science and Technology of Japan.

Figure captions

Fig. 1. Cross section of fabricated pseudo spin-valve MTJ (a) and exchange biased spin valve MTJ (b).

Fig. 2. Magnetoresistance loops of pseudo spin-valve $Co_{40}Fe_{40}B_{20}/MgO/Co_{40}Fe_{40}B_{20}$ MTJs (a) and $Co_{20}Fe_{60}B_{20}/MgO/Co_{20}Fe_{60}B_{20}$ MTJs (b) at room temperature (open circles) and 5 K (solid circles). MTJs were annealed at 450°C. TMR ratios are as high as 450% at RT and 747% at 5 K for MTJ with $Co_{40}Fe_{40}B_{20}$ electrodes and 472% at RT and 804% art 5 K for one with $Co_{20}Fe_{60}B_{20}$ electrodes. (b) Temperature dependence of resistance (open circles) and TMR ratio (solid circles) in parallel and anti-parallel configurations in the $Co_{40}Fe_{40}B_{20}/MgO/Co_{40}Fe_{40}B_{20}$ MTJ.

Fig. 3. (a)TMR ratios as a function of $T_a$ for P-SV MTJs (open circles) and EB-SV MTJs (solid circles). (b) Exchange bias field as a function of $T_a$ for EB-SV MTJs. Insets show the TMR loops of an EB-SV MTJ annealed at 350°C (top) and 450°C (bottom).

Fig. 4. TEM images in regions of CoFeB/MgO/CoFeB interface for P-SV MTJs annealed at 450°C (a), EB-SV MTJs annealed at 375°C (b), and EB-SV MTJs annealed

at 450ºC (c).

Fig. 5. EDX spectrum profiles of points shown in Fig. 4 (b) and (c) for EB-SV MTJs annealed at 375ºC ((a)-(c)) and 450ºC ((d)-(f)).

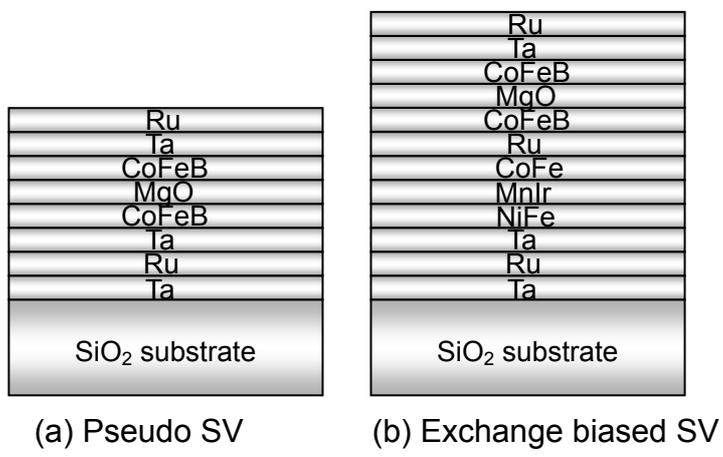

Fig. 1   Hayakawa et al

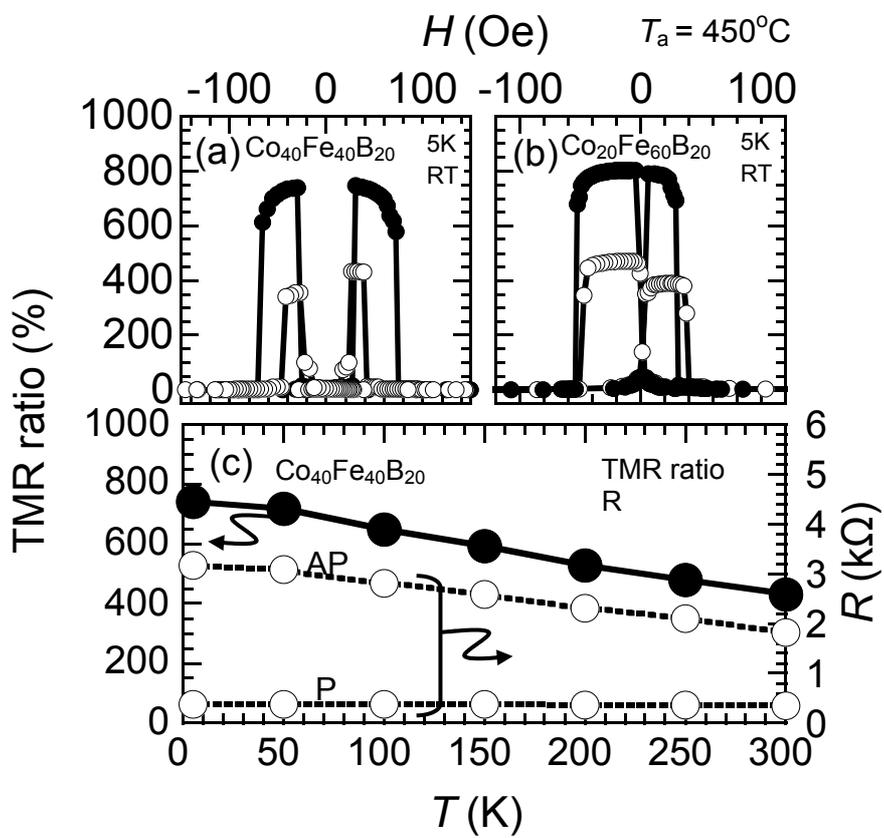

Fig. 2  Hayakawa et al

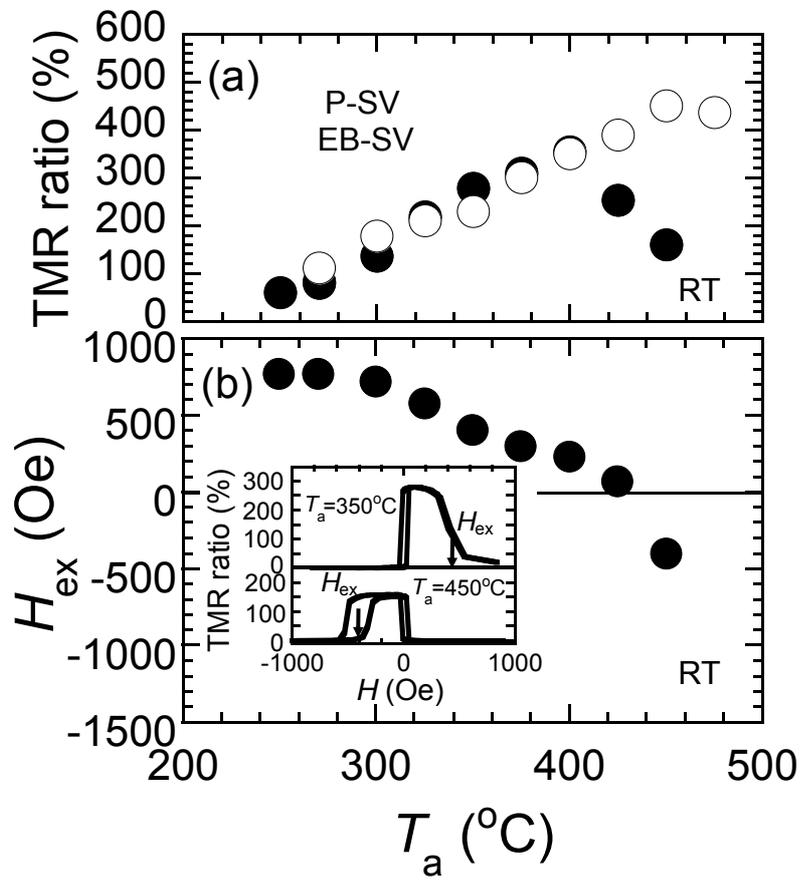

Fig. 3  Hayakawa et al

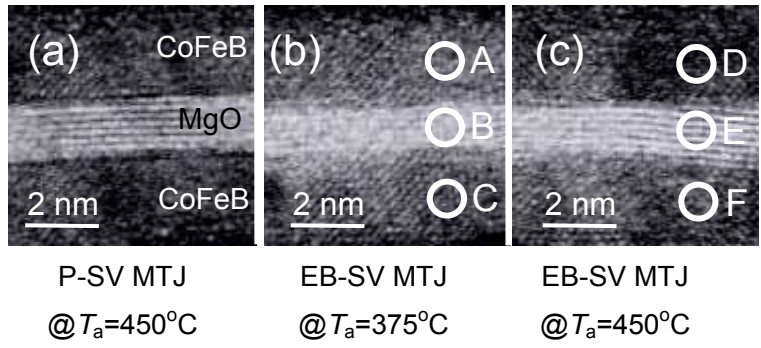

Fig. 4 Hayakawa et al

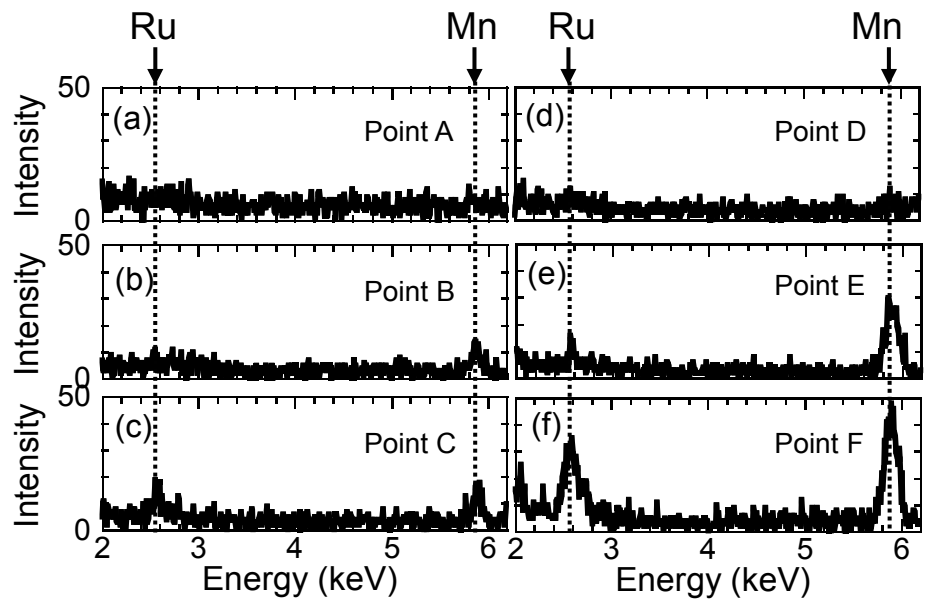

Fig. 5  Hayakawa et al